\documentclass[12pt]{article}

\usepackage{amsmath,amssymb}
\usepackage{amsfonts}

\renewcommand{\baselinestretch}{1.1}
\begin{document}

\newcommand{\be}{\begin{equation}}
\newcommand{\ee}{\end{equation}}
\newcommand{\bea}{\begin{eqnarray}}
\newcommand{\eea}{\end{eqnarray}}
\newcommand{\ack}[1]{[{\bf Pfft!: #1}]}
\newcommand{\CP}[1]{{\mathbb{CP}}^{#1}}
\newcommand{\C}{\mathbb{C}}
\newcommand{\Diff}{{\rm Diff}}
\newcommand{\Gr}{{\rm Gr}}
\newcommand{\Xtr}{X_{\rm tr}}
\newcommand{\Xlong}{X_{\rm long}}
\newcommand{\Seff}{S_{\rm eff}}
\newcommand{\msusy}{m_{\rm susy}}

\newcommand{\eref}[1]{(\ref{#1})}
\def\NPB{{\it Nucl. Phys. }{\bf B}}
\def\PL{{\it Phys. Lett. }}
\def\PRL{{\it Phys. Rev. Lett. }}
\def\PRD{{\it Phys. Rev. }{\bf D}}
\def\CQG{{\it Class. Quantum Grav. }}
\def\JMP{{\it J. Math. Phys. }}
\def\SJNP{{\it Sov. J. Nucl. Phys. }}
\def\SPJ{{\it Sov. Phys. J. }}
\def\JETPL{{\it JETP Lett. }}
\def\TMP{{\it Theor. Math. Phys. }}
\def\IJMPA{{\it Int. J. Mod. Phys. }{\bf A}}
\def\MPL{{\it Mod. Phys. Lett. }}
\def\CMP{{\it Commun. Math. Phys. }}
\def\AP{{\it Ann. Phys. }}
\def\PR{{\it Phys. Rep. }}

\hyphenation{Min-kow-ski}
\hyphenation{cosmo-logical}
\hyphenation{holo-graphy}
\hyphenation{super-symmetry}
\hyphenation{super-symmetric}

\renewcommand{\baselinestretch}{1.5}

\rightline{gr-qc/0406037}
\rightline{VPI-IPPAP-04-05}
\centerline{\Large \bf }\vskip0.25cm
\centerline{\Large \bf }\vskip0.25cm
\centerline{\Large \bf Toward a Background Independent}
\vskip0.15cm
\centerline{\Large \bf Quantum Theory of Gravity}
\vskip 1cm

\renewcommand{\baselinestretch}{1.1}

\renewcommand{\thefootnote}{\fnsymbol{footnote}}
\centerline{{\bf Vishnu Jejjala,\footnote{{vishnu@vt.edu}}
Djordje Minic,\footnote{{dminic@vt.edu}}
Chia-Hsiung Tze\footnote{{kahong@vt.edu}}
}}
\vskip .5cm
\centerline{\it Institute for Particle Physics and Astrophysics}
\centerline{\it Physics Department, Virginia Tech}
\centerline{\it Blacksburg, VA 24061, U.S.A.}
\vskip .5cm

\begin{abstract}
Any canonical quantum theory can be understood to arise from the compatibility
of the statistical geometry of distinguishable observations with the canonical
Poisson structure of Hamiltonian dynamics.
This geometric perspective offers a novel, background independent
non-perturbative formulation of quantum gravity.
We invoke a quantum version of the equivalence principle, which requires both
the statistical and symplectic geometries of canonical quantum theory to be
fully dynamical quantities.
Our approach sheds new light on such basic issues of quantum gravity as the
nature of observables, the problem of time, and the physics of the vacuum.
In particular, the observed numerical smallness of the cosmological 
constant can be rationalized in this approach.

\end{abstract}

\setcounter{footnote}{0}
\renewcommand{\thefootnote}{\arabic{footnote}}

\newpage

\noindent
Quantum theory and the General Theory of Relativity constitute the pillars
of modern physics.
The difficulty in reconciling the predictions of each with the other
signals the fundamental incompleteness of our understanding of Nature.
Local quantum field theories, which incorporate the demands of the Special
Theory of Relativity into the structure of quantum mechanics, are
spectacularly predictive.
Incorporating the demands of the General Theory of Relativity into the
structure of quantum mechanics should proffer new insights about space,
time, matter, and dynamics.

At first glance, a perturbative approach to quantum gravity, such as string
theory, assimilates a canonical and pragmatic quantum theory of measurement
with the underlying universal physics of the gravitational interaction.
By providing a way to compute boundary observables (a generalized $S$-matrix),
string theory serves as a natural candidate for the {\it definition} of
quantum gravity in asymptotically infinite spaces \cite{banks}.
Yet even its most successful avatar, non-perturbative string theory
on asymptotically anti-de Sitter ($AdS$) backgrounds \cite{ads},
cannot address the local (or more precisely, quasi-local) questions
associated with observers in particular regions of spacetime.
More severely, string theory does not as yet provide a background independent
formulation of a quantum theory of gravity.
In order to compute physically interesting quantities (transition
amplitudes, for instance), additional data about the structure of the 
asymptopia are needed.
Indeed, it is still unclear whether it is even reasonable to expect a
background independent formulation for theories built out of 
``$S$-matrix-like''
observables.

Other paths to quantum gravity \cite{ashtekar} have their own obstacles,
for example,
defining the meaning of time and time evolution,
determining the observables appropriate to a background independent 
formulation of physics,
recovering the Standard Model, which is a self-consistent theory of chiral 
matter, and
accounting for the presence of structures at large and small scales, a fact
seemingly irreconcilable with the full diffeomorphism invariance of the 
theory.

Previous constructions rest on the foundation of canonical quantum theory.
Therefore, one expects, by hypothesis, to end up with a structure 
buttressed by the
scaffolding of ordinary quantum mechanics.
Inspired both by the conceptual and technical challenges of quantum gravity as
well as by exciting new data about our cosmological
background \cite{WMAP}, we have recently proposed a different architecture 
for quantum gravity \cite{tzeminic}.

Our analysis
is motivated by the conceptual tension in the way results of
measurements are treated in quantum mechanics and in general relativity
and by the stark contrast between the rigid geometry of canonical quantum
theory and the dynamical geometry of spacetime.
In the $\hbar\to 0$ limit, there is a correspondence principle that recovers
classical structures.
Working {\it ab initio} from a generalized quantum theory,
we seek a suitable limit in which quantum gravity becomes general relativity
in a manifest and observable way.

It is well known that standard quantum mechanics is fully captured by the 
geometry
of $\CP{n}$ \cite{tzeminic, geomqm}, a homogeneous, isotropic, and simply
connected K\"ahler manifold with constant, holomorphic sectional curvature.
The rigid structure of $\CP{n}$ welds the statistical geometry of
distinguishable observations with the canonical Poisson structure of
Hamiltonian dynamics.
All the main features of quantum mechanics are embodied in $\CP{n}$. Thus,
the superposition principle is tied to viewing $\CP{n}$ as a collection of
complex lines passing through the origin.
Entanglement arises from the embeddings of the products of two complex
projective spaces in a higher dimensional one. The
geometric phase stems from the symplectic structure of $\CP{n}$.
Furthermore, there is a deep interconnection between the geometric 
properties of
quantum mechanics and the geometry of spacetime.
In the $\hbar\to 0$ limit, the Fisher--Fubini--Study metric on the quantum
phase space described by $\CP{n}$ reduces to a spatial metric provided 
that the
configuration space for the quantum system under consideration {\it is} 
physical space itself.
Similarly, the time parameter of the evolution equation can be related to the
quantum metric via $\hbar\,ds = \Delta E\,dt$, where
the dispersion in energy for a given Hamiltonian is determined by $\Delta E$.
Finally, the Schr\"odinger equation is the geodesic equation for an abstract
``particle'' of Yang--Mills charge
moving on $\CP{n} = {U(n+1)}/{(U(n) \times U(1))}$ in the presence of an
effective external field (namely, the $U(n) \times U(1)$ valued curvature
two-form), whose source is the Hamiltonian \cite{tzeminic, geomqm}.

The above geometric structure of canonical quantum mechanics is beautifully
tested by experiment and is highly robust from the geometric point of view.
Unlike various past generalizations,
our proposal \cite{tzeminic} builds gravity into the very foundation of
an extended quantum mechanical theory.
It does so by making the kinematical structure compatible with the generalized
dynamical structure.
{\it The quantum symplectic and metric structure, and therefore the almost
complex structure, are themselves fully dynamical.}
The underlying physical reason for this more general dynamical framework
develops from a quantum version of the equivalence principle, which provides a
foundational underpinning for a non-perturbative quantum theory of gravity.

The intuition behind the classical equivalence principle is to demand
indistinguishability of a local gravitational field with acceleration.
The intuition behind the quantum equivalence principle is to demand the
validity of canonical quantum mechanics in every local neighborhood, that is
to say, {\it in the space of quantum events}.
Vector fields on a manifold are described locally on tangent spaces, and the
dynamics are then patched together.
Similarly, we envision a larger geometric structure whose tangent spaces
are the canonical Hilbert spaces of quantum mechanics.
The tangent spatial transverse metric emerges from the quantum
metric, by assuming that the underlying configuration space is space itself.
The transverse space is a classical moduli space, the space of
inequivalent degenerate vacua of the theory.
Time is a measure of the geodesic distance in this general space of
statistical events. The nature of time is hence probabilistic.
The longitudinal spatial coordinate corresponds to the dimensionality of
the tangent Hilbert space.

When the metric and symplectic form on phase space become fully
dynamical entities,
only individual quantum events make sense observationally.
Physics is required to be diffeomorphism invariant in the sense of information
geometry provided that the statistical metric and the symplectic structures
remain compatible. This condition requires a {\it strictly} ({\it i.e.}\
non-integrable) almost complex structure on the generalized space of quantum
events.
This extended framework readily implies that the wavefunctions labeling the
event space, while still unobservable, are in fact irrelevant.
They are as meaningless as coordinates in general relativity.
The physics does not rely on such choices.
At the basic level, there are only dynamical correlations of
quantum events, and the observables are furnished by diffeomorphism invariant
quantities in the quantum configuration space.
Time evolution is contained in the relational properties of general
information metrics.
The concept of a vacuum state becomes meaningful only in the presence of an
emergent spacetime background consistent with the rigid geometry of the
canonical quantum theory.

To find the kinematical arena for this generalized framework, one seeks a
coset of $\Diff(\C^{n+1})$ that locally looks like $\CP{n}$ and also allows
for mutually compatible metric and symplectic structures, expressed in the
existence of a (generally non-integrable) almost complex structure.
The nonlinear Grassmannian
$\Gr(\C^{n+1}) = \Diff(\C^{n+1})/\Diff(\C^{n+1},\C^n \times \{0\})$,
with $n \to \infty$ fulfills these requirements \cite{vizman}.
This space is a natural generalization of $\CP{n}$.
The logic here is exactly as in general relativity.
Just as we gauge the Lorentz symmetry into the general
diffeomorphism group, the
Grassmannian is a gauged version of complex projective space, which is the 
geometric
realization of quantum mechanics.
As with standard geometric quantum mechanics,
the geodesic length on the space of events assigns probabilities.
The non-integrability of the almost complex structure translates to the
existence only of a {\it local} notion of time and a local metric on the 
space of
quantum events.
The temporal evolution equation locally is the geodesic equation:
${d u^a}/{d\tau} + \Gamma^{a}_{bc} u^b u^c =
\frac{1}{2 E_p}Tr(H F^a_b) u^b$, where
$H$ is the Hamiltonian,
$\Gamma^{a}_{bc}$ is the affine connection associated with this general 
metric $g_{ab}$, and
$F_{ab}$ is the $\Diff(\C^{n+1},\C^n \times \{0\})$ valued curvature two-form.
The parameter $\tau$ is defined via $\hbar\, d\tau = 2 E_p\, dt$,
with $E_p$ the Planck energy.
(This is consistent with the energy-time uncertainty relation.)
The geodesic equation follows from the conservation of the energy-momentum
tensor, $\nabla_a T^{ab} = 0$.
Since both the metric and symplectic data are contained in $H$ and are $\hbar
\to 0$ limits of their quantum counterparts, we have a consistent nonlinear
``bootstrap'' between the space of quantum events and the generator of the
dynamics \cite{tzeminic}.

The diffeomorphism invariance of the new quantum phase space is explicitly
taken into account in the following dynamical scheme:
\begin{equation}
\label{BIQM1}
R_{ab} - \frac{1}{2} g_{ab} R  - \lambda g_{ab}= T_{ab}
\end{equation}
($\lambda = \frac{n+1}{\hbar}$ for $\CP{n}$; in that case $E_p \to \infty$).
This is the Einstein--Yang--Mills equation on the space of quantum events.
Moreover, we must demand for compatibility
\begin{equation}
\label{BIQM2}
\nabla_a F^{ab} = \frac{1}{E_p} H u^b.
\end{equation}
The two equations imply, via the Bianchi identity, a conserved
energy-momentum tensor, which
together with the conserved ``current''
$j^a \equiv \frac{1}{2E_p} H u^a$, $\nabla_a j^a =0$,
results in the generalized geodesic Schr\"{o}dinger equation.
As in general relativity, it will be crucial to understand both the local and
global features of various solutions to the dynamical equations.
The kinematical structure of ordinary quantum mechanics is compatible with our
general dynamical formulation, and thus we expect that our formalism is
compatible with all known cases in which quantum theories of gravity have been
non-perturbatively defined, albeit in {\it fixed} asymptotic backgrounds (such
as string theory in asymptotically $AdS$ spaces) \cite{vishnu}.
Practical calculational aspects can be related to toy models for deforming the
symplectic structure both in the classical and quantum regimes \cite{tatsu}.

What determines the actual form(s) of the Hamiltonian?
The only requirement is that $H$ should define a canonical quantum mechanical
system whose configuration space is space and whose dynamics define a
consistent quantum gravity in a flat background.
We are aware of only {\it one} example satisfying this criterion:
{\it Matrix theory} \cite{matrix}.
Thus, our proposal defines a background independent, non-perturbative,
holographic formulation of M-theory \cite{tzeminic}.
At first sight, the choice of Hamiltonian may appear {\it ad hoc} given the
generality of our scheme.
However, through the geodesic form of the Schr\"{o}dinger equation,
$H$ should be viewed as a ``charge'' and is thus determined in a
quantum theory of gravity through the non-trivial topology of the quantum
phase space \cite{geometro}.
This may well be the case because the nonlinear Grassmannian, which is the 
space of events,
is not simply connected \cite{vizman}.
One important element of this new approach to quantum gravity
is the existence of a correspondence limit between the background independent
quantum theory (with a Matrix theory Hamiltonian) and the classical background
independent Einstein theory of gravity coupled to matter.
This connection is ensured by the emergent property of spacetime in a
background independent way.

Finally, let us see how the above approach sheds new light on reconciling the
numerical smallness of the vacuum energy (cosmological constant) with respect
to the Planck scale \cite{vishnu}.
In our framework, the {\it mean value} of the vacuum energy density is
identically {\it zero} given the diffeomorphism symmetry in the space of
quantum events.
Locally, the vacuum energy is determined by the quantum theory in the tangent
space, which {\it by construction} is Matrix theory.
Unbroken supersymmetry in Matrix theory ensures that the vacuum energy 
vanishes identically within a given local neighborhood.
The obstruction to extending this into a global statement arises from patching
together the physics of the tangent spaces at different points in the space of
probabilities.
The observed value of the cosmological constant has a natural 
interpretation as
a {\it fluctuation} about the zero mean.
It signals the dynamical (cosmological) breaking of supersymmetry.

The magnitude of the fluctuation in the cosmological constant can be 
estimated by recalling that in the limit when the classical background 
emerges, we must also recover the standard canonical quantum fluctuations 
of measurable quantities.
The central limit theorem yields the Fisher metric
as the unique homogeneous, isotropic, and simply connected metric on
the space of distinguishable statistical events.
The hypotheses of the central limit theorem require that the number of quantum
events be large.
Thus, consistent with the law of large numbers, fluctuations obey a
Poisson distribution.
The fluctuation (measured value) of the cosmological constant $\Delta \Lambda$
is related to its conjugate quantity, the volume of spacetime $\Delta V$,
through the uncertainty principle:
$\Delta \Lambda \, \Delta V \sim \hbar$.
As the number of quantum events scales as the spacetime volume,
$\Delta V \sim \sqrt{V}$;
this in turn implies (in natural units) that
\be
\Delta \Lambda \sim \frac{1}{\sqrt{V}},
\label{cc}
\ee
a result consistent with the observed value for the cosmological constant
given the observed volume of spacetime \cite{WMAP, sorkin}.

How should we motivate a spacetime volume large enough to accommodate the
smallness of the observed vacuum energy?
How does Planck scale physics relate to dynamics in the infrared?
The spacetime uncertainty relation \cite{ur} provides a compelling 
explanation.
In perturbative string theory,
modular invariance on the worldsheet translates in target space to
the spacetime uncertainty relation:
$\Delta T \, \Delta \Xtr \simeq \ell_s^2 \simeq \alpha'$.
(Here, $T$ is a timelike direction;
$\Xtr$ is a spacelike direction transverse to the lightcone.)
Nonperturbatively, the eleven dimensional Planck length $\ell_p = 
M_p^{-1}$ and the size of the
M-theory circle that extends at large string coupling sets the scale 
$\alpha' = \ell_p^3/R$.
The radius $R$ determines the maximal uncertainty in $\Xlong$, the 
longitudinal
direction in the Matrix theory limit, which implies that
$\Delta T \, \Delta \Xtr \, \Delta \Xlong \simeq \ell_p^3$,
a cubic relation consistent with the existence of membranous structures
in M-theory \cite{ur}.
The Planck energy and the geodesic distance on the probability space are 
related
by $\hbar\, \Delta s = 2 E_p\, \Delta T$.
We estimate the line element on
the space of probabilities to scale as $ds \simeq e^{-\Seff}$, where 
$\Seff$ denotes the low-energy (Euclidean)
effective action for the matter degrees of freedom propagating in an 
emergent (fixed) spacetime background.
The spacetime uncertainty relation becomes
\be
\Delta \Xtr \, \Delta \Xlong \simeq e^{\Seff}\, \ell_p^2.
\ee
The product of the ultraviolet cutoff (the maximal uncertainty in the 
transverse coordinate)
and the infrared cutoff (the maximal uncertainty in the longitudinal 
coordinate)
is thus exponentially suppressed compared to the Planck scale.
There is a {\em gravitational see-saw} \cite{seesaw}.
Within an emergent spacetime background, at low energies, we locally have 
a supergravity theory,
but there is in general no globally defined supersymmetry.
The effective scale of cosmological supersymmetry breaking $\msusy$ 
provides an estimate for the vacuum energy as implied by the 
gravitational see-saw:
\be
\omega^4 \simeq \left( \frac{\msusy^2}{M_p} \right)^4.
\ee
The cosmological breaking of supersymmetry argues the intimate connection
between the volume of spacetime and the observed vacuum energy \cite{seesaw}.
In the $\omega \to 0$ limit, we recover supersymmetry in spacetime and
diffeomorphism invariance in the space of quantum events.
The numerical smallness of the vacuum energy density is thus consistent with
the principle of naturalness in that the vanishing of a dimensionful quantity
restores a dynamically broken symmetry.

\section*{Acknowledgments}
\noindent 
The work of VJ and DM is supported in part by the U.S.\ Department of
Energy under contract DE-FG05-92ER40677.

\end{document}